\documentstyle[aps,prl,preprint]{revtex} 
\draft  
 
\begin{document} 
\author{Xiao-Hua Wu$^{1,2}$,Hong-Shi Zong$^1$,Hou-Rong Pang$^1$,Fan Wang$^1$} 
\address{$^1$,Department of Physics and Center for Theoretical Physics,  
Nanjing University, Nanjing 210093, China}
\address{$^2$ Department of Physics,Sichuan Union University,Chengdu 610064,China}
\title{A New Bell Inequality for Two Spin-1 Particle System }
\maketitle 
  
\begin{abstract} 
For a two spin-1 particles system, we  derive a new Bell's type inequality 
for local hidden variables model. For the 
singlet state for two spin-1 particles, we show that the inequality is 
violated while it is satisfied for the direct product state.

E-mail: HXW@chenwang.nju.edu.cn.   
\end{abstract} 
 

  In 1965, Bell demonstrated that an interpretation of quantum theory
 in terms of local hidden-variables (LHV) is impossible[1], using 
inequalities now universally known as Bell inequalities. In 1969,
Clauser ${\sl et}$ ${\sl  al.}$ [2] showed that these inequalities might be test exprimentally. Since then, a series of experiments, of increasing precision and making
use of various atomic sources and detection arrangments, have been carried out
to test one version or another  of Bell inequalities [3,4]. These inequalities studied are based on a $2\otimes2$ Hilbert space. For a particle,
the case that it's dimension is more than 2 has been discussed by the famous
Bell-KS theorem[5], which concerns the results of a (counterfactual) set of measurements on quantum state described by a vector in a three dimensional Hilbert 
space. They consider, for example, measurements of the squares of the three
angular momentum components of a spin-1 state. They assume that the corresponding operators commute and can be measured simultaneously, providing one "yes" and two "no's" to
the  questions "Does the spin componet along $\hat{a}$,$ \hat{b}$,
$\hat{a}\times\hat{b}$ vanish?" for any $\hat{a}\perp\hat{b}\in S^2$, the unit
sphere in ${\bf R}^3$. Specker [6] and Bell [7] observed that Gleason's 
theorem [8] implies that there can be no assignment of "yes's" and "no's" to
the vector of $S^2$ consistent with this requirement: each triad is "colored" with one "yes" and two "no's". Certainly, Bell-KS theorem can be viewed  as a proof for the contradictions between noncontextual LHV models and quantum mechanics.

However, there are several pionts which weakened  the Bell-KS theorem:
first, noncontextual LHV is just a special case of a more general one discussed in Bell's inequalities for $2\otimes2$ Hilbert space; second, the Bell-KS theorem does not depend on the entanglement of the state,  a spine -1 particle's state is enough, while the Bell's inequalities is not violated by a direct product state[9]. In other words, the contradiction of LHV models and quantum mechanics 
should depend on the fact that the system is entangled or not; and lastly, there are recently
 works [10-13] pointed out that finite precision measurement will nullify the
Bell-KS theorem, this make the Bell-KS theorem untestable since the fact that 
 the measurements yet known have finite precision. Comparing with studying of the  spin-half particles, the situation of the higher spin case is rather unsatisfying. Recently, M.Zukowski ${\sl et}$ ${\sl al.}$ [14] showed that higher-dimensional two-particle entanglements are realizable via multiport beam splitters, and the results presented in their paper move the discussion on entangled higher-than-1/2 spin systems from the realm of gedanken-experiments to real expriments. Being motivated by their work, here we shall let the $3\otimes3$ case to be considered in a way similiar to what Bell's theorem has done for the $2\otimes2$ case.
There has been a series of works studying the case that the Hilbert dimension N for each particle is more than two. First results, in 1980-1982, suggested that the conflict between local realism and quantum mechanics diminishes with growing N [15-17]. 
In the early 1990's Peres and Gisin [18-19] considered certain dichotomic observables applied to maximally entangled pairs of particles, they showed that the violation of local realism survives, while N is growing, but never exceeds the factor $\sqrt{2}$. Recently, D.Kaszlikowski
${\sl et} $ ${\sl al}$ [20] investigated the general case of two entangled quantum systems defined in N-dimensional Hilbert spaces( they called it "quNits"), and via a numerical linear optimization method they showed that violations of local realism are stronger for two maximally entangled quNits
$( 3\le N \le 9)$ than that for two quNits and they are increase with N , while the two quNit systems is described by a special mixed state. In present paper,only the case N=3 is concerned:
 first we shall derive a inequality for a $3\otimes 3$ Hilbert space from locality and reality; then, using the singlet state of two spin-1 particles system,
we shall show that the inequality is violated. The inequality is satisfied for
the direct product state of the two spin-1 particles system.

The singlet state for two spin-1 particles are
\begin{eqnarray}
\vert \Psi\rangle=\frac{1}{\sqrt{3}}(\vert m_1=1 \rangle \vert m_2=-1\rangle
-\vert m_1=0\rangle \vert m_2=0 \rangle
+\vert m_1=-1\rangle \vert m_2=1 \rangle),
\end{eqnarray}
where $\vert m_i\rangle$ denotes the eigenvetor of spin operator $\hat{S}$ along the
direction z, $\hat{S}_i(z)\vert m_i \rangle =m_i \vert m_i\rangle$,
$(m_i=1,0,-1)$ for particle i(i=1,2). Let $\vert m'_i\rangle$ to be the eigenvector of $\hat{S}(\beta_i)$, $\hat{S}(\beta_i)\vert m'_i\rangle =m'_i\vert m'_i\rangle$( for simplity, we have let the directions  in the x-z plane, and each direction is viewed as a rotation $\beta$ along the y axis), there is a connection between $\vert m_i\rangle$ and $\vert m'_i\rangle $
\begin{eqnarray}
\vert m'_i\rangle=\Sigma^3_{j=1} D_{ji}(\beta)\vert m_j\rangle,
\end{eqnarray}

and the rotation matrix is 

\begin{eqnarray}
D(\beta)=\left| \begin{array}{ccc}
                    \frac{1+cos(\beta)}{2}&\frac{sin(\beta)}{\sqrt{2}}&
                       \frac{1-cos(\beta)}{2} \\
                    \frac{sin(\beta)}{\sqrt{2}}&cos(\beta)&-\frac{1-cos(\beta)}{\sqrt{2}} \\
\frac{1-cos(\beta)}{2}& \frac{sin(\beta)}{\sqrt{2}}& \frac{1+cos(\beta)}{2} 
                    \end{array} \right|
\end{eqnarray}

With the singlet state(1) as a source, particle 1 propagates along the y axis, while particle 2 in the -y axis. A Stern-Gerlach magnetic analyzer is put in a place where particle 1 will arrive at, and it will give the results for spin projection along a direction $\beta_1$, while a similar analyzer, which locates a distance away from the analyzer for particle 1, is used to measure the spin projection along $\beta_2$ for particle 2. Now, the singlet state (1) can be written as

\begin{eqnarray}
\vert \Psi \rangle =\frac{1}{\sqrt{3}}
\{ 
&&\sin^2(\frac{\beta_1-\beta_2}{2})\vert 1\rangle\vert 1\rangle
-\frac{1}{\sqrt{2}}\sin(\beta_1-\beta_2)\vert 1 \rangle \vert 0\rangle
+\cos^2(\frac{\beta_1-\beta_2}{2})\vert 1\rangle \vert -1\rangle
\nonumber \\
&&+\frac{1}{\sqrt{2}}\sin(\beta_1-\beta_2)\vert 0\rangle \vert 1 \rangle 
-\cos(\beta_1-\beta_2)\vert 0\rangle \vert 0 \rangle 
-\frac{1}{\sqrt{2}}\sin(\beta_1-\beta_2)\vert 0\rangle \vert -1 \rangle
\\ \nonumber
&&+\cos^2(\frac{\beta_1-\beta_2}{2})\vert -1\rangle \vert 1 \rangle
+\frac{1}{\sqrt{2}}\sin(\beta_1-\beta_2) \vert -1 \rangle \vert 0 \rangle 
+\sin^2(\frac{\beta_1-\beta_2}{2})\vert -1\rangle \vert -1\rangle\}.
\end{eqnarray}

Defining the joint probability correlation
\begin{eqnarray}
P_{m_1m_2}=\langle\Psi\vert m_1\rangle\langle m_1\vert
\otimes \vert  m_2\rangle\langle m_2\vert \Psi \rangle,(m_1,m_2=1,0,-1)
\end{eqnarray}
the singlet state in the form (4) give the following joint probabilities:
\begin{eqnarray}
P_{11}(\beta_1,\beta_2)=\frac{1}{3}\sin^4(\frac{\beta_1-\beta_2}{2}),
P_{00}+P_{0,-1}+P_{-1,0}+P_{-1-1}=\frac{1}{3}
[1+\sin^4(\frac{\beta_1-\beta_2}{2})].
\end{eqnarray}
We should prove that the correlation (6) can not be interpreted by LHV models.

Assuming that the state(1) can be described by a set of parameters $\lambda$,
LHV models gives the probabilities $p_m(\beta
_1,\lambda)$ and
$q_n(\beta_2,\lambda)$ for the two results that the spin projection along 
$\beta_1$ is m for particle 1 and the spin projection along $\beta_2$ is n for 
particle 2 respectively( m,n=1,0,-1).
 As a cosequence of the relation
\begin{eqnarray}
\vert m_i=1\rangle \langle m_i=1\vert
+\vert m_i=0\rangle\langle m_i=0 \vert
+\vert m_i=-1 \rangle \langle m_i=-1\vert=I, (i=1,2)
\end{eqnarray}
there is a natural conditions here 
\begin{eqnarray}
&&o\le p_1(\beta_1,\lambda)
+p_0(\beta_1,\lambda)
+p_{-1}(\beta_1,\lambda)\le 1, \nonumber \\
&&0\le q_1(\beta_2,\lambda)
+q_0(\beta_2,\lambda)
+q_{-1}(\beta_2,\lambda)\le 1.
\end{eqnarray}

It should be noted that $p_m(\beta_1,\lambda)$ does not depend on
the settings of $\beta_2$ , while $q_n(\beta_2,\lambda)$ is not depending on 
$\beta_1$  either. They are required by
the locality assumption, which aserts that experiments done on one place have
no influence on the resluts of measurement done on the other place located a distance away 
if two measurements are performed simultaneously, and the joint probability should be
\begin{eqnarray}
P_{mn}(\beta_1,\beta_2)
=\int_{\lambda \in \Lambda} p_m(\beta_1,\lambda)
q_n(\beta_2,\lambda) \rho(\lambda) d\lambda,
\end{eqnarray}
while 
\begin{eqnarray}
\int_{\lambda\in\Lambda}\rho(\lambda)d\lambda=1.
\end{eqnarray}

 In order to derive a inequality, we use the following simple algebraic 
theorem: giving six real numbers x,x',X,y,y' and Y, such that 
 $0\leq{x},x'\leq{X}, 0\leq{y},y'\leq{Y}$
one must always have 

\begin{eqnarray}
-XY{\leq}xy-xy'+x'y+x'y'-x'Y-Xy{\leq}0
\end{eqnarray}
The proof of it has been given by Clauser and Horne [21]. Making the 
identifications
$x=p_1(\beta_1,\lambda), x'=p_1(\beta'_1,\lambda),
y=q_1(\beta_2,\lambda), y'=q_1(\beta'_2,\lambda)$,
and taking X=Y=1(since the natural conditions), one can obtain
\begin{eqnarray}
&&p_1(\beta_1,\lambda)q_1(\beta_2,\lambda)
-p_1(\beta_1,\lambda)q_1(\beta_2',\lambda)
+p_1(\beta'_1,\lambda)q_1(\beta'_2,\lambda)
\nonumber \\
&&+(p_0(\beta'_1,\lambda)+p_{-1}(\beta'_1,\lambda))
(q_o(\beta_2,\lambda)+q_{-1}(\beta_2,\lambda))\le 1,
\end{eqnarray}
through using the natural conditions$(8)$. Intergrating over $\rho(\lambda)$,
a inequality for $3\otimes 3$ Hilbert space can be dereived :
\begin{eqnarray}
&&S=P_{11}(\beta_1,\beta_2)
-P_{11}(\beta_1,\beta'_2)
+P_{11}(\beta'_1, \beta'_2)
\nonumber \\
&& +P_{00}(\beta'_1,\beta_2)
+P_{0-1}(\beta'_1,\beta_2)
\\ \nonumber
&&+P_{-10}(\beta'_1,\beta_2)
+P_{-1-1}(\beta'_1,\beta_2)\le 1.
\end{eqnarray}
It can be easily shown tha this inequality will be violated by the quantum joint probability given in eq.(6). Choosing the following sets of the angles
$\beta_1=0$,$\beta'_1=2\beta_2$, $\beta'_2=3\beta_2$, and $\beta_2=147.7$degree,
we get a contradiction
\begin{eqnarray}
S=1.12\le 1.
\end{eqnarray}
Certainly, the above S is less than the factor 4/3 given in case of the multiport beam splitter[14].

If the state is a direct product state, for example
\begin{eqnarray}
\vert \Psi'\rangle=\vert 1\rangle\vert 0\rangle, 
\end{eqnarray}
now the probability is 
\begin{eqnarray}
&&P_{11}(\beta_1,\beta_2)=\cos^4\frac{\beta_1}{2}\sin^4\frac{\beta_2}{2},
P_{00}=\frac{1}{4}\sin^2\beta_1\sin^2\beta_2,
P_{0,-1}=\frac{1}{2}\sin^2\beta_1\cos^4\frac{\beta_2}{2},\\ \nonumber
&&P_{-1,0}=\frac{1}{2}\cos^4\frac{\beta_1}{2}\sin^2\beta_2,
P_{-1,-1}=\sin^4\frac{\beta_1}{2}\cos^4\frac{\beta_2}{2},
\end{eqnarray}
then, we have the following form
\begin{eqnarray}
S=\cos^4\frac{\beta_1}{2}\sin^4\frac{\beta_2}{2}
-\cos^4\frac{\beta_1}{2}\sin^4\frac{\beta'_2}{2}
+\cos^4\frac{\beta'_1}{2}\sin^4\frac{\beta'_2}{2}
+(1-\cos^4\frac{\beta'_1}{2})(1-\sin^4\frac{\beta_2}{2}),
 \end{eqnarray}
while the ineqality (11) can be transfered into a form
\begin{eqnarray}
xy-xy'+x'y'+(1-x')(1-y)\le1.
\end{eqnarray}
Comparing the above two forms, we know that $ S \le1 $ is always satisfied for
direct product state and has no relation  to the directions  chosen.

In conclusion, for the  $3\otimes3$ Hilbert space, we have derived a inequality as 
Bell's inequality for the $2\otimes2$ case. For the  singlet state for two spin-1 particles system, we show that the inequality will be violated while it is valid for the case when the two particle in a direct product state.

This work is supported by the NSF,SED,SSTD of China.

\end{document}